\documentclass[fleqn,10pt]{wlscirep}
\usepackage{threeparttable}
\usepackage{float}

\title{Measuring Economic Activities of China with Mobile Big Data}

\author[1,2]{Lei Dong}
\author[1]{Sicong Chen}
\author[1]{Yunsheng Cheng}
\author[1]{Zhengwei Wu}
\author[1]{Chao Li}
\author[1,*]{Haishan Wu}
\affil[1]{Big Data Lab, Baidu Research, Baidu, Beijing, 100085, China}
\affil[2]{School of Architecture, Tsinghua University, Beijing, 100084, China}

\affil[*]{wuhaishan@baidu.com}

\begin{abstract}

Emerging trends in smartphones, online maps, social media, and the resulting geo-located data, provide opportunities to collect traces of people's socio-economical activities in a much more granular and direct fashion, triggering a revolution in empirical research. These vast mobile data offer new perspectives and approaches for measurements of economic dynamics and are broadening the research fields of social science and economics. In this paper, we explore the potential of using mobile big data for measuring economic activities of China. Firstly, We build indices for gauging employment and consumer trends based on billions of geo-positioning data. Secondly, we advance the estimation of store offline foot traffic via location search data derived from Baidu Maps, which is then applied to predict revenues of Apple in China and detect box-office fraud accurately. Thirdly, we construct consumption indicators to track the trends of various industries in service sector, which are verified by several existing indicators. To the best of our knowledge, we are the first to measure the second largest economy by mining such unprecedentedly large scale and fine granular spatial-temporal data. Our research provides new approaches and insights on measuring economic activities.

\end{abstract}
\begin{document}

\flushbottom
\maketitle

\section*{Introduction}

Mobile internet, especially location-aware services, are ubiquitous in our everyday life: each time when we open an application (APP), search a nearby restaurant, check the route and traffic, take a car using ride-hailing APP, use the mobile map navigation, user’s location is sensed via various positioning technologies, generating massive mobility trace data. Such kind of big data, which directly reflects user’s social and economic behaviors, provides new tools to measure the dynamics of economy in a real-time fashion, overcoming the limitation of timeliness and sample size of traditional survey methodology, and thus making deep influence in empirical research [\cite{einav2014economics,varian2014big, lazer2009computational}].

China, the second largest economy in the world with about 600 million smartphone users [\cite{website:smartphone}], has being profoundly impacted by mobile internet, and meantime it is struggling to transform its economy from investment-led to consumer-driven. Thus the indicators or indices to gauge China’s economic activities are very valuable for entrepreneurs, researchers, investors, and most importantly, policy makers. However, there are three primary challenges when addressing this data-driven issue. First, government statistics are generally released with a lag of weeks or months, and for some large-scale surveys (e.g. economic census), the coarse aggregate data would only be available several years later. It is undoubted that more timely indicators could be helpful to make more timely decisions for both governments and companies. Second, many researchers question the reliability and quality of official data. For example, Gross Domestic Product (GDP) data is considered to be overstated [\cite{rawski2001happening}] and (registered) unemployment rate is suspected to be understated as it remains steady at 4\%, hardly impacted by economic slowdown [\cite{feng2015long}]. Moreover, many important economic indicators, such as surveyed unemployment, are not publicly available, casting a veil over the measurement of China’s economic activities. Last but most important, the changing economic structure raises new challenges for measuring the emergence of service industries like the retails, restaurants, entertainment, finance, etc., which are increasingly making up a considerable proportion of the whole economy but difficult to quantify previously. 

Facing above-mentioned challenges, researchers previously resorted to new data sources such as search queries [\cite{ettredge2005using, askitas2009google, choi2009predicting, choi2012predicting, scott2013bayesian, goel2010predicting, preis2013quantifying, yang2015forecasting}], social media [\cite{antenucci2014using, llorente2015social, bollen2011twitter, asur2010predicting}], satellite images [\cite{chen2011using, henderson2012measuring, michalopoulos2013pre, mellander2015night-time, website:spaceknow}], online commodity price [\cite{cavallo2015scraped, cavallo2013online}], financial transactions [\cite{agarwal2014consumption, gelman2014harnessing, website:morgan, website:unionpay}], check-in data[\cite{website:foursquare}] and mobile phone data [\cite{deville2014dynamic, blumenstock2015predicting, toole2015tracking, smith2016beyond}] etc., to build socio-economical indicators or study economic behaviors from different perspectives. 

Varian and coauthors show the possibility of using Google search query indices for short-term economic prediction [\cite{choi2009predicting,choi2012predicting, scott2013bayesian}]. And web search data is also proofed to be helpful in forecasting consumer behavior [\cite{goel2010predicting}], and even financial market [\cite{preis2013quantifying}]. Similar methodology is applied to social media, researchers use Twitter data to create indices to predict unemployment [\cite{antenucci2014using, llorente2015social}], stock market [\cite{bollen2011twitter}], and box-office earnings [\cite{asur2010predicting}], to name a few. 

However, analysis of search queries or social media texts is arguable, especially when handling the words with different meanings [\cite{choi2012predicting}]. Additionally, these works rely on 'ground truth' or official statistics as baseline models, and then improve the forecasting performance by utilizing search query or social media text data. 

Another line of thought seeks proxy variables to measure economic activities directly, especially for countries in which official statistics are not available or reliable. For example, night light data is widely used in measuring economic output [\cite{chen2011using, henderson2012measuring, michalopoulos2013pre}] and poverty [\cite{xie2015transfer}]. And online product price is also crawled by researchers for constructing online price index, which can not only replicate government figures closely in the United States, but also helpful for countries lacking trustful statistics [\cite{cavallo2013online,cavallo2015scraped}]. In industry, SpaceKnow and Baidu use satellite images and online web data to create Satellite Manufacturing Index [\cite{website:spaceknow}] and Small Businesses Indices [\cite{website:baidu}] to gauge China's economic activities, respectively.

Recently, more and more individual level data, especially financial transactions and mobile phone data are used by academia and industries to study human's economic activities. For transactions data, Gelman et al. [\cite{gelman2014harnessing}] have measured people's response of spending to anticipated income with 75,000 user's transaction data captured by Check (a financial service application) in the United States. On contrary, Agarwal and Qian [\cite{agarwal2014consumption}] study how consumers respond to unanticipated income shock by bank transactions of 180,000 individuals in Singapore. China UnionPay also develops economic indicators from the expenditures of bank card in various market segments [\cite{website:unionpay}]. For mobile phone data, Toole et al. [\cite{toole2015tracking}] track employment shocks using mobile phone Call Detail Records (CDRs) in Europe, and Blumenstock et al. [\cite{blumenstock2015predicting}] infer poverty and wealth at individual level by combining Rwanda's mobile phone metadata and survey data with machine learning algorithms. 

In fact, although quite useful, mobile phone CDRs or metadata are still an indirect way to measure the economy, which need to be combined with survey or official statistics to train models. And the spatial resolution of mobile phone data at cell tower level ranges from hundreds of meters in the urban core to thousands of meters in rural areas, making it difficult to track firm level economic activities. 

The prevalence of smartphones and the generated geo-located data and mobility trace data allows us to measure the economic activities of China in a much more direct fashion at more granular level compared with other data sources that have been explored. In this paper, we build \textbf{Employment Index} and \textbf{Consumer Index} to measure the trends of employment in industrial parks and consumers in commercial areas by using billions of geo-positioning points. We then propose models to estimate consumer foot traffic volumes of store locations in service sector using location search and navigation data derived from Baidu Maps. We first apply it to predict revenues of Apple retail stores and box-office earnings in China, and achieve satisfying results. We then construct \textbf{Consumption Trends Index} to track consumer spending trends of various industries (e.g. auto sales, restaurant, financial investment, tourism, etc.) in service sector.

To the best of our knowledge, we are the first to measure the second largest economy by mining such unprecedentedly large scale and fine granular spatial-temporal data. Our research, providing new insights into understanding China’s economy, is not designed to supplant traditional statistics based survey, but to supplement such indicators, if reliable, to achieve more abundant measurements.

\section*{Data Description and Privacy}
\subsection*{Data Sources}
Four datasets are used in this study: 

1) Geo-positioning data. As the largest online map service provider in China, Baidu Maps provides location search and positioning services for hundred of millions users, generating tens of billions of location requests everyday. Each location point includes anonymized ID, the coordinate (longitude and latitude, respectively) and timestamp. In our research, geo-positioning data starts from January 2014 to June 2016. The spatial distribution of positioning data is shown in Figure 1. 

\begin{figure}[ht]
\centering
\includegraphics[width=\linewidth]{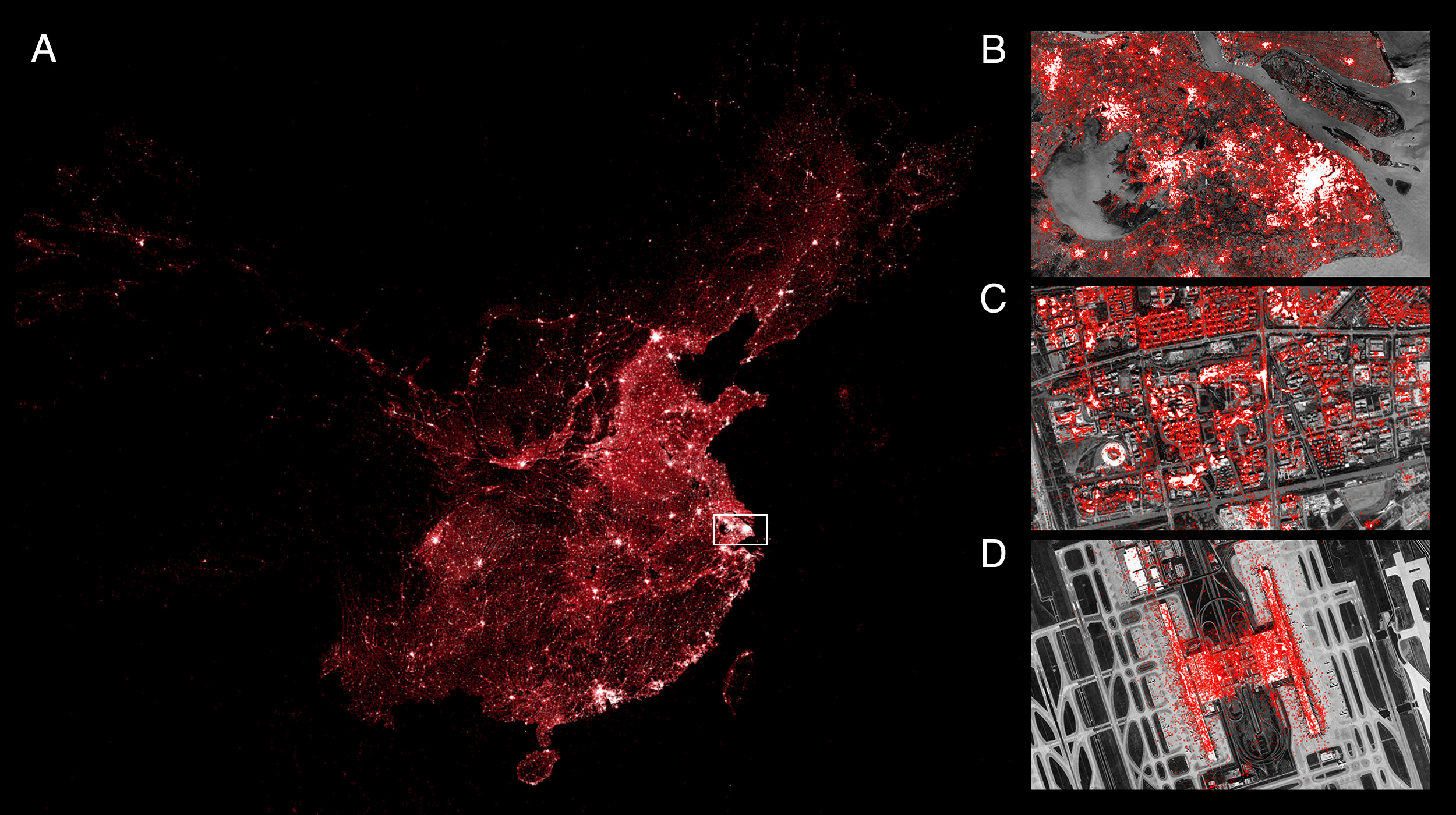}
\caption{Spatial-temporal big data reflects human activities at different scales. (A) At national level, data points image the fact that most of China's population is concentrated in large cities in the east. The brighter the spot, the more data points (population) are aggregated. (B) At regional level, figure shows urban clusters in the Yangtze River Delta. (C) At zone level, Zhangjiang Hi-tech Park in Shanghai. (D) At building level, Pudong Airport in Shanghai. }
\label{fig:figure1}
\end{figure}

2) Location search data from Baidu Maps. Each location search data (i.e. map query) includes anonymized IDs, query keywords, the returned POI ID, and timestamp. Location search behavior is strongly correlated with user’s actual visitation to the queried POI, and therefore as detailed in the Results, we use online map query data to estimate offline foot traffic. Map query data also starts from January 2014 to June 2016. 

3) Points of Interest (POIs) data, corresponding to a specific location such as restaurant, hotel, shopping mall etc. The whole dataset comprises of about 50 million POIs, all of which are classified into different categories by machine learning algorithms, for example , `Walmart’ belongs to 'Supermarket’ category which is a subcategory of `Shopping’ (POI category is detailed in the Supplementary Materials). A map query action could be regarded as a consumer demand or future check-in for a certain kind of place or category that the issued location belongs to.

4) Areas of Interest (AOIs) data, corresponding to a specific region such as industrial parks, commercial areas, scenic regions, etc. In order to measure the economic activities of employment and consumer, we manually labeled about 6,000 AOIs, including 2,000 large industrial parks and 4,000 commercial areas. The spatial distribution of these areas is shown in Figure 2.

\begin{figure}[H]
\centering
\includegraphics[width=0.9\textwidth]{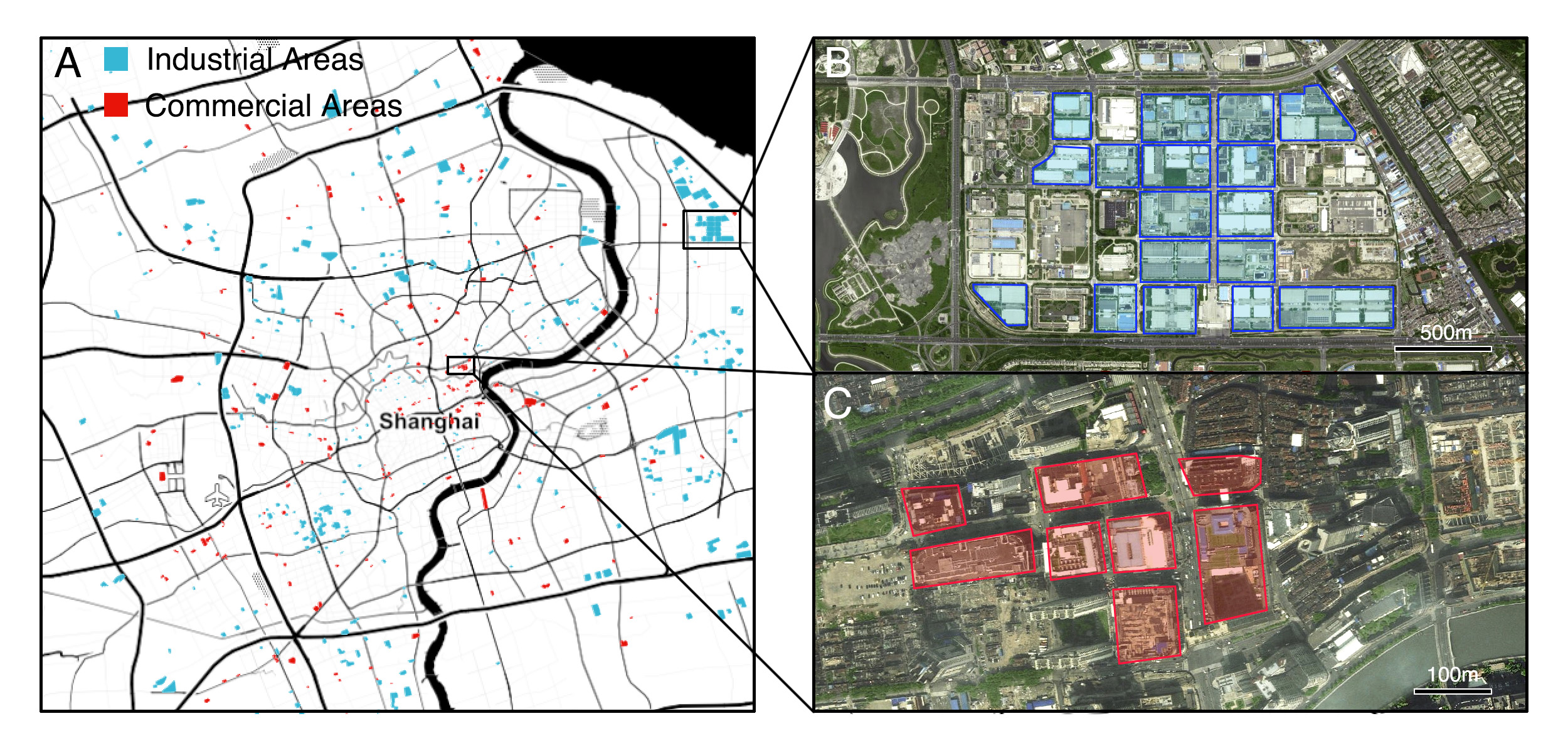}
\caption{(A) Illustration of the spatial distribution of labeled AOIs. Industrial areas are colored in blue and commercial areas are in red. Zoomed-in part: (B) and (C) are industrial parks and commercial centers, respectively .}
\label{fig:figure2}
\end{figure}

\subsection*{Data Privacy}
We adopt very rigorous protocol to protect user’s privacy in this research: 
\begin{itemize}
\item All user IDs in our data are hashed and anonymized to ensure that one cannot associate the data to individual users.
\item All the data are saved in secure servers and can only be accessed with strict procedures by high standards. 
\item All the researchers must follow a confidential agreement to use data for approved research. We solely focus on aggregated instead individual level to measure the economic activity in this research.
\end{itemize}

\section*{Methodology}

We here focus on measuring the trends and constructing indicators of employment and consumption, two most essential elements related to economic activities. We reduce this task to count the number of employees and consumers appear in specific location, such as industrial parks, shopping malls, attractions, and cinemas etc. by analyzing mobility traces derived from geo-positioning data and location search query data. We then track their growth trends, monitor the volatilities to evaluate related economic performance.

Intuitively, it seems not difficult to identify and sum the employees and consumers, however, this task is very technically challenging in implementation because:

1) Accurately identifying employees and consumers associated to one site, namely a POI in our data, is very difficult solely by POIs and mobility trace data, largely due to the fact that POIs simply include location coordinates without spatial boundary information. What's more, many consumer sites, e.g. movie theaters, supermarkets, restaurants, etc., are located together in large buildings, which makes it difficult to tell them apart from mobility traces. Though some machine learning models have been proposed in computer science community to correspond POIs to mobility traces (named trajectory semantic problem) and applied in several business scenarios, even the state-of-the art models [\cite{furletti2013inferring}] are not sufficiently accurate for rigorous social and economic science research.

2) When mining the trends and constructing indicators from any type of online data collected passively, especially mobile location data, search queries, and social media texts etc., the changing users' structure is an unavoidable issue. And the adjustment of provided online services and the following dynamics of the data may also lead to bias [\cite{ruths2014social,lazer2014parable}].

To overcome these difficulties and obtain relatively reliable economic measurements, we employ the following strategies:

Firstly, we label about 6,000 AOIs, of which 2,000 are industrial parks and 4,000 are commercial areas. Compared with POIs, the labeled AOIs, with obvious layout boundaries and specific land use, are more reliable and robust to associate with employees and consumers. We then identify the employees and consumers appeared in those AOIs from their mobility traces.

%==should we hide some details to file the patent?
Secondly, like traditional survey practices which carefully draw samples based on various criteria, we sample mobile users who consistently have location points during a 13-month rolling window (at least one point each month). In other words, the results of any month is based on the `continuous users' of the past 13-month. By this way, we are able to draw a stable and sufficient large sampled users, and ensure these users will not be affected by the adjustment of provide online services, resulting in more robust measurements. And, the 13-month long window also provides a convenient way to calculate year-over-year changes of figures.

Thirdly, in order to track the trends of consumer spending of various consumption industries in service sector, we estimate the volume of foot traffic of specific consumer sites (e.g. retails, cinemas, auto sales, restaurants) by corresponding locations search volumes (map queries). As shown in our previous research [\cite{zhou2016early, xu2016store}], searching a location on maps leads to high probability of offline arrival in the near future, and on the aggregate level, we find that the the volume of map queries of one location is highly correlated with its offline foot traffic. We therefore use the volume of map queries to estimate the volume of visited consumers, and define the trends of location queries of different industries as consumer spending indicators.

\section*{Results}

\subsection*{1. Employment and Consumer Indices}

Employment and consumption are two major indicators for economic activities. At the micro level, the changes of the number of employees may reflect the performance of one specific company: if the business of one company grows, more jobs would be created to expand production or services; on the contrary, the pace of layoffs may accelerate when it faces difficulties from business operation or weakening market. These phenomena are bound up with the macro-economy closely. For example, when companies suffered serious crisis in the beginning of 2009, China's economy also experienced rapid decline with exports stumbling and demand contraction, leading to ever-increasing unemployment pressure.

\subsubsection*{1.1 Tracking employment change through location data}

We here assume that the numerical changes of employees and consumers could be tracked from geo-positioning data, by which we propose to construct employment and consumer indices to measure China's economic activities. In order to verify this idea, we study four cases with obvious employment volatility: two of which announced mass layoffs, and the remaining two experienced rapid developments with employment surge. We first identify employees and count the monthly number in these areas. Each result is normalized by the value of the first month. 
%去掉了see the Methods部分

\begin{figure}[ht]
\centering
\includegraphics[width=\linewidth]{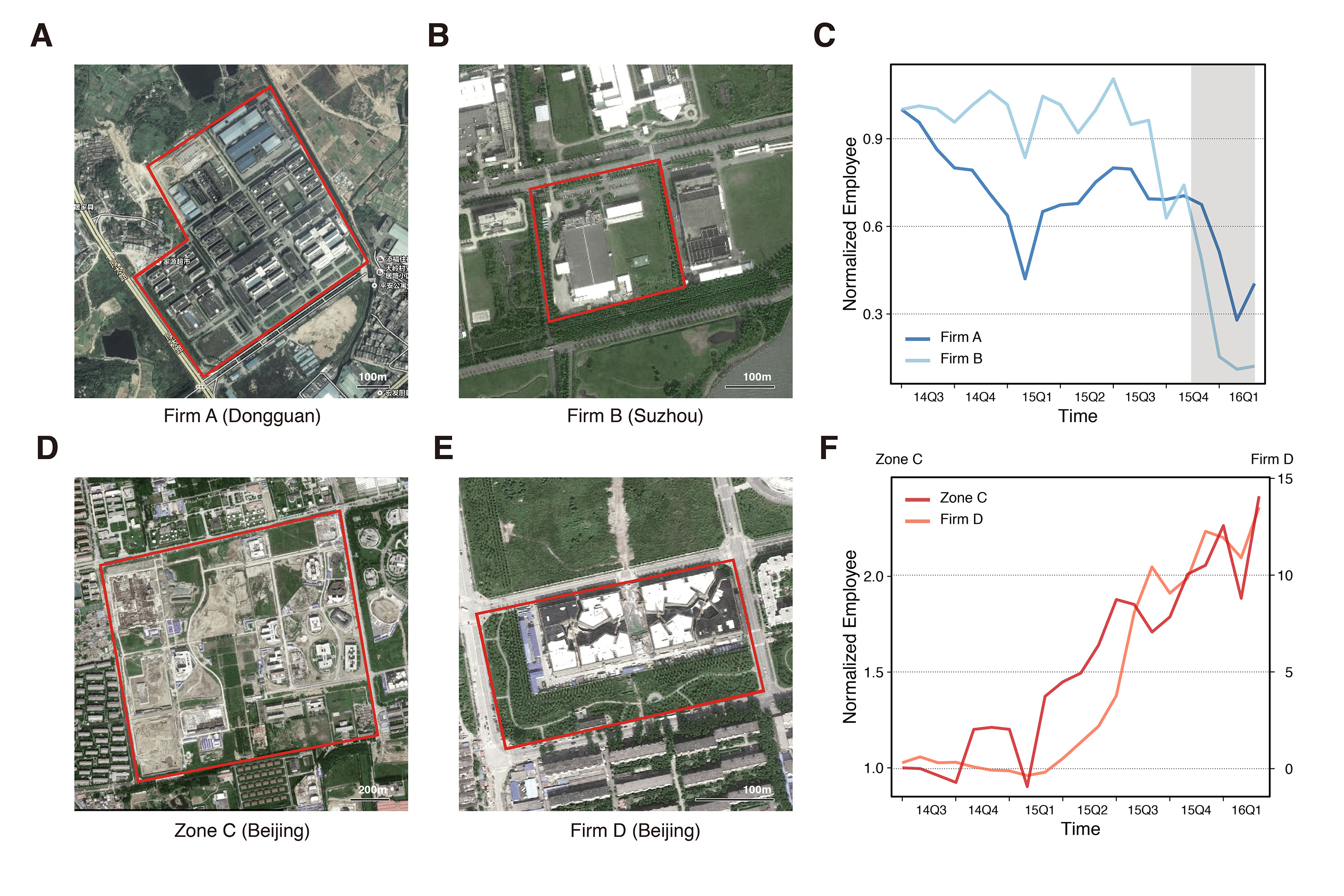}
\caption{Measure employment changes at company/zone level. We study four typical cases, Firm I (A) and Firm II (B) announced mass layoffs, and Zone III (D) and Firm IV (E) have been experiencing rapid growth. We compute the monthly number of employees in these areas and the normalized data is shown in (C) and (F). The closing time of Firm I and II is marked by gray bar in (C).}
\label{fig:figure3}
\end{figure}

Figure 3A-C show the first two cases. Firm I is a large shoe factory located in Dongguan, a southeast coastal city of Guangdong Province, where is the heart of the country's export manufacturing. Due to global industrial transfer, increasing labor cost and reduced export order, the plant was closed down at the first quarter of 2016, leaving thousands of workers jobless. As shown in Figure 3C, there is a sharp drop in the number of employees (the dark blue line) during this period (gray box). A similar shock hit Firm II, a mobile phone factory located in Suzhou of Jiangsu Province (Figure 3B). It was shut down at the end of December 2015 with hundreds of workers losing their jobs. It is worth noting that Firm I gradually laid off the employees over the past three years while Firm II shut down the whole factory suddenly at the end of 2015. The discrepancy between these two cases is also mirrored in our results as demonstrated in Figure 4C: the employment of Firm I slipped 20\% in 2015 than in 2014, with a consequent decline of 30\% year-over-year in the first quarter of 2016, while the employment of Firm II remained stable and then dipped dramatically at the last quarter of 2015 after the close down.

Figure 3D-F show the second two cases of increasing employment. Zone III is a software park in Beijing (Figure 3D) with lots of high-tech companies. As most of them started to relocate there from 2014, and since then, as shown in Figure 3F (red line), the number of employees has almost doubled. Firm IV is a fast growing start-up based in Beijing (Figure 3E) with rapid expansion since 2015. As the figure shows, its employment jumped in the second quarter of 2015 after it raised billions of investment funding.

\subsubsection*{1.2 Employment Index}
%Revised by Haishan 20160711
We therefore build employment index to track the macro trends of employment by aggregating employment changes of 2,000 industrial parks consisting of 1000 traditional industrial parks (mainly manufacturing industry) and 1,000 high-tech (mainly information technology, bio-tech etc.) industrial parks. Figure 4 illustrates the monthly employment index from January 2014 to June 2016, in which gray, blue and red line indicate the employment index of all industrial parks, traditional industrial parks, and high-tech industrial parks, respectively. All indices are normalized by the mean value of year 2014 which is set to be 100. 

%Since macro-economic activities result from the aggregation of micro-level behaviors, we merge the data of 2,000 industrial areas. For further analysis, we split 2,000 industrial areas into three groups, gray line is the overall trends including all 2,000 areas, blue line only includes about 1,000 areas of manufacturing industry, and red line contains areas except for manufacturing, mainly software, bio-tech, and logistics parks (Figure 4). All figures are normalized by the mean value of year 2014 with the base number of 100. 

%Revised by Haishan 20160711
One should notice that we did not perform seasonal adjustment on current index given that robust adjustment generally requires more than three-year long time series while ours cover two year and half at present. Besides, Spring Festival holiday (Chinese New Year), one of most important holiday in China, is set by lunar calendar and falls on different dates between January and February each year. This holiday officially lasts seven days, but many workers may take several weeks off, leading dramatic economic slow during this time of period. To reduce its effect, we aggregate the employment changes of January and February each year together to calculate the year-over-year growth rate. 

%One should notice that we don’t apply any seasonal adjustment method to figures, which is usually done for removing the seasonal component of time series data and analyzing the underlying trends. Since our datasets cover less than three years so far, there is insufficient time-series to do seasonal adjustment. Besides, the timing of the long lunar Spring Festival holiday also makes it difficult to adjust Chinese data in January and February. The holiday officially lasts seven days, but many workers take several weeks off, economic activities slow dramatically around the holiday, which is set by the lunar calendar and falls on a different time period in the first quarter of every year. Thus, we aggregate the data of January and February together to reduce the influence of Spring Festival holiday when calculating the year-over-year numbers (Figure 4B).

Figure 4A shows the monthly Employment Index of industrial parks without seasonal adjustment since January 2014, which exhibit seasonal volatility of the index, and most obviously, the sharp drop during Spring Festival of each year. Figure 4B shows the corresponding year-over-year growth rate. Our figures show that: (1) The overall employment index of industrial parks was relatively steady with slow decline over last two years and started falling from the beginning of 2016, indicating the overall economy slowdown. Specifically, it dropped by 3\% in May 2016 and slightly rose in June on a year-over-year basis. (2) The monthly employment index of manufacturing parks shows a sustained and more obvious decline than overall performance, such downward trend is similar in official manufacturing purchasing mangers index (PMI)[http://data.stats.gov.cn]. It dipped more than 5\% year-over-year over most time of period, indicating that this sector is seriously suffering from the economic slowdown. (3) The employment index of high-tech parks, by contrast, saw a healthy and continuous increase, with more than 3.5\% year-over-year over the last two years. Despite the similar decline since January 2016, it started growing slowly from June 2016, suggesting that high-tech companies from software and bio-tech industries, which refers to as the `new economy', remain creating more jobs than other sectors. 

\begin{figure}[ht]
\centering
\includegraphics[width=\linewidth]{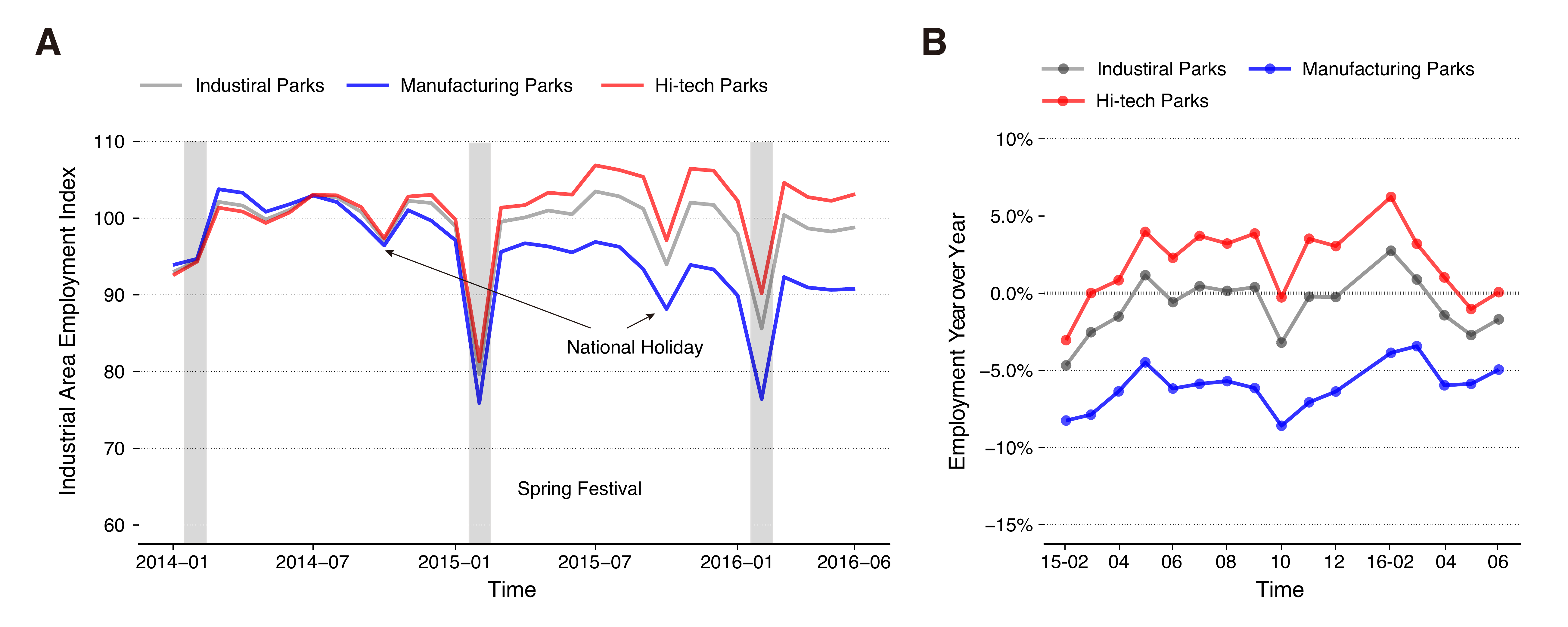}
\caption{Employment Index of industrial parks. (A) Gray line is the overall trend of all 2,000 industrial parks which is relatively steady with slow decline; blue line is the result of about 1,000 manufacturing industrial parks which shows a sustained and more obvious downward trend; red line is the result of about 1000 hi-tech parks (e.g. software, bio-tech) which presents a healthy and continuous growth. All figures are normalized by the mean value of year 2014 which is set to be 100, and Spring Festival is marked by gray bar. (B) Monthly year over year growth rate. January and February are aggregated together to remove the influence of Spring Festival holiday, which falls on a different time period in January or February every year.}
\label{fig:figure4}
\end{figure}

\subsubsection*{1.3 Consumer Index}
%Revised by Haishan 20160711
By the similar fashion, we track the changes of consumer from about 4,000 major commercial areas covering most of the big shopping malls in China. We then construct monthly Consumer Index to measure the change and track the growth of consumers in these areas. Figure 5A and 5B illustrate our consumer index and corresponding year-over-year growth rate, respectively. The consumer index generally climbs up during summer holiday (July and August) and before Spring Festival (December and January), and goes down during Spring Festive (February). One can see that the Consumer Index is gradually inching down, and its year-over-year growth rate also fluctuates with more dips below zero than spikes over time, indicating the weakening consumer demand possibly dragged down by the overall slowdown and online e-commerce. The year-over-year growth rate of Consumer Index once started rising since September 2015 and peaked at around 9\% in December, however, from the beginning of 2016, it fell back below zero and sank to near -7\% in June this year, indicating the relatively low consumption demand in the first half of 2016.  

\begin{figure}[H]
\centering
\includegraphics[width=\linewidth]{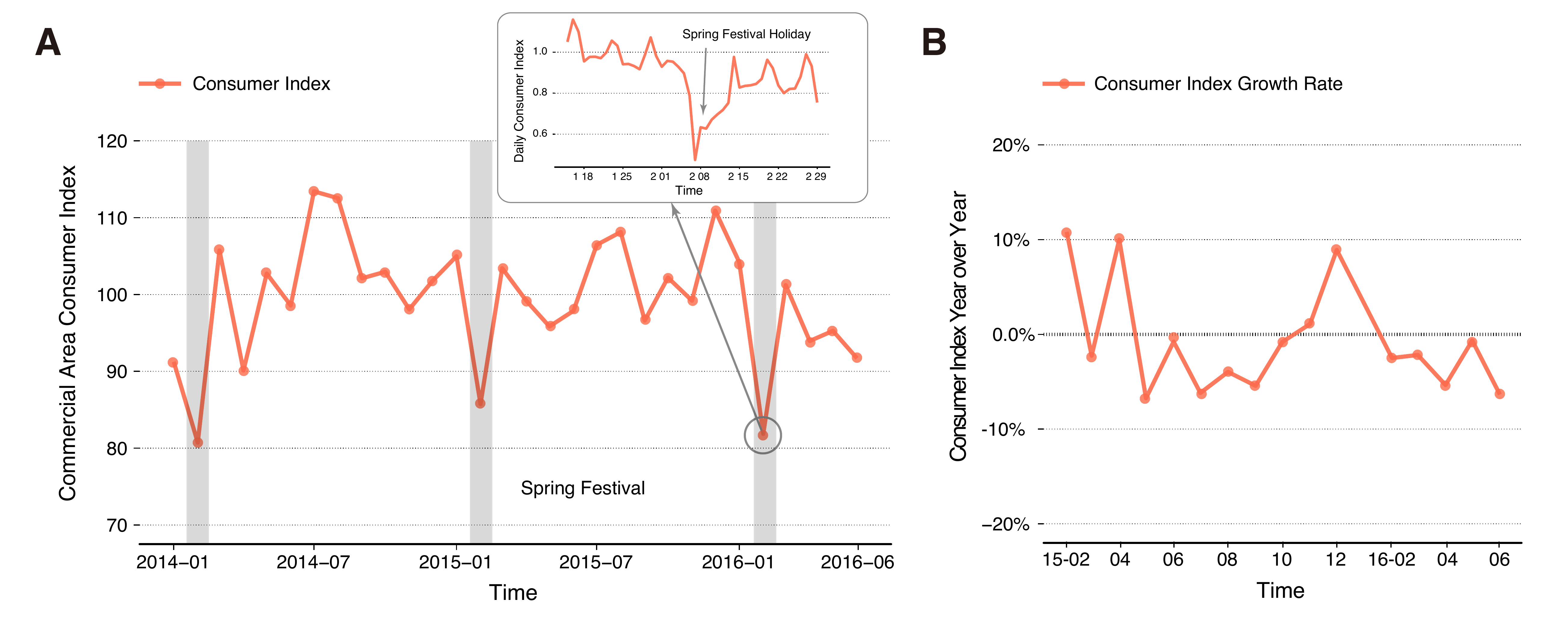}
\caption{Consumer Index of commercial areas. (A) Orange line is the monthly Consumer Index, which is normalized by the mean value of year 2014 with the value of 100, and Spring Festival is marked by gray bar. (B) Monthly year over year growth rate. The consumer index generally climbs up during summer holiday (July and August) and near Spring Festival (December and January), and goes down during Spring Festive (February). The year-over-year growth rate fluctuates with more dips below zero than spikes over time, showing that the trend of consumer index is going down.}
\label{fig:figure5}
\end{figure}

The proposed Employment Index and Consumer Index measures the economic performance of selected industrial parks and commercial areas in China, pictures the complicated reality of China's economy, and provides us a new perspective to chart the macro-economic performance from a bottom-up view.

\subsection*{2. Foot Traffic Estimation, Revenue Forecast, and Consumption Trends Index}

In the previous section, we have construct the consumer index by counting the number of visited consumers in large commercial areas via geo-positioning dataset, we then seek to track the consumer spending trends of various industries in service sector. This task, however, is challenging because, as described in the Methodology section, identifying user's offline visitation of specific POIs, such as a restaurant or a retail shop instead of an AOI such as a shopping center, by solely using geo-location data cannot achieve ideal accuracy. We here resort to a new data source, location search query data from Baidu Maps (i.e. map query data), to address the following issues: (1) Is it possible to estimate the offline foot traffic of one specific location? (2) Can we `nowcast' or even forecast the consumer spending or revenue of one location or a firm with chain stores? (3) How can we build consumer spending indices to evaluate the performance of different industries in service sector?

\subsubsection*{2.1 Foot Traffic Estimation from Map Query Data}

Foot traffic is one of the most critical metrics for service sectors such as retails, restaurants, movies, and hotels, just to name a few. We observe that, with no surprise, the volume of map queries regarding one specific location is highly correlated with that of offline foot traffic. Such correlation is illustrated in Figure 6, which shows the number of people searching for Shopping Mall I and Shopping Mall II (orange dash line) on Baidu Maps and the foot traffic (red solid line) of each place (estimated by geo-positioning dataset). In both cases map queries and foot traffic present obvious periodical patterns, peaking on weekends and dipping during weekdays. To evaluate the efficiency and accuracy of foot traffic estimation via map query data, we compared two seasonal auto regression (AR) models to predict the foot traffic:

\begin{equation}\label{eq:1}
y_t = \beta_{0} + \beta_{1} y_{t-1} + \beta_{2} y_{t-7} + e_{t}
\end{equation}
\begin{equation}\label{eq:2}
y_t = \beta_{0} + \beta_{1} y_{t-1} + \beta_{2} y_{t-7} + \beta_{3} q_{t} + e_{t}
\end{equation}

where $y_t$, $y_{t-1}$, and $y_{t-7}$ are the numbers of consumers at time $t$, $t-1$, and $t-7$, respectively. $q_t$ is the volume of map queries and $e_t$ is the error term. The regression results shown in Table 1 indicate that adding map queries in regression improves in-sample fit significantly. The $R^{2}s$ of Shopping Mall I and II increase from 0.727, 0.767 to 0.914, 0.931, respectively. Actually, if we just use map queries to fit the foot traffic, the results are surprisingly accurate: the $R^{2}s$ are 0.880 for Shopping Mall I and 0.910 for Shopping Mall II (Figure 6CF). Therefore, we could use map queries to estimate the volume of foot traffic and evaluate the performance of foot-traffic-dominated stores, firms and industries in service sectors.

\begin{figure}[ht]
\centering
\includegraphics[width=\linewidth]{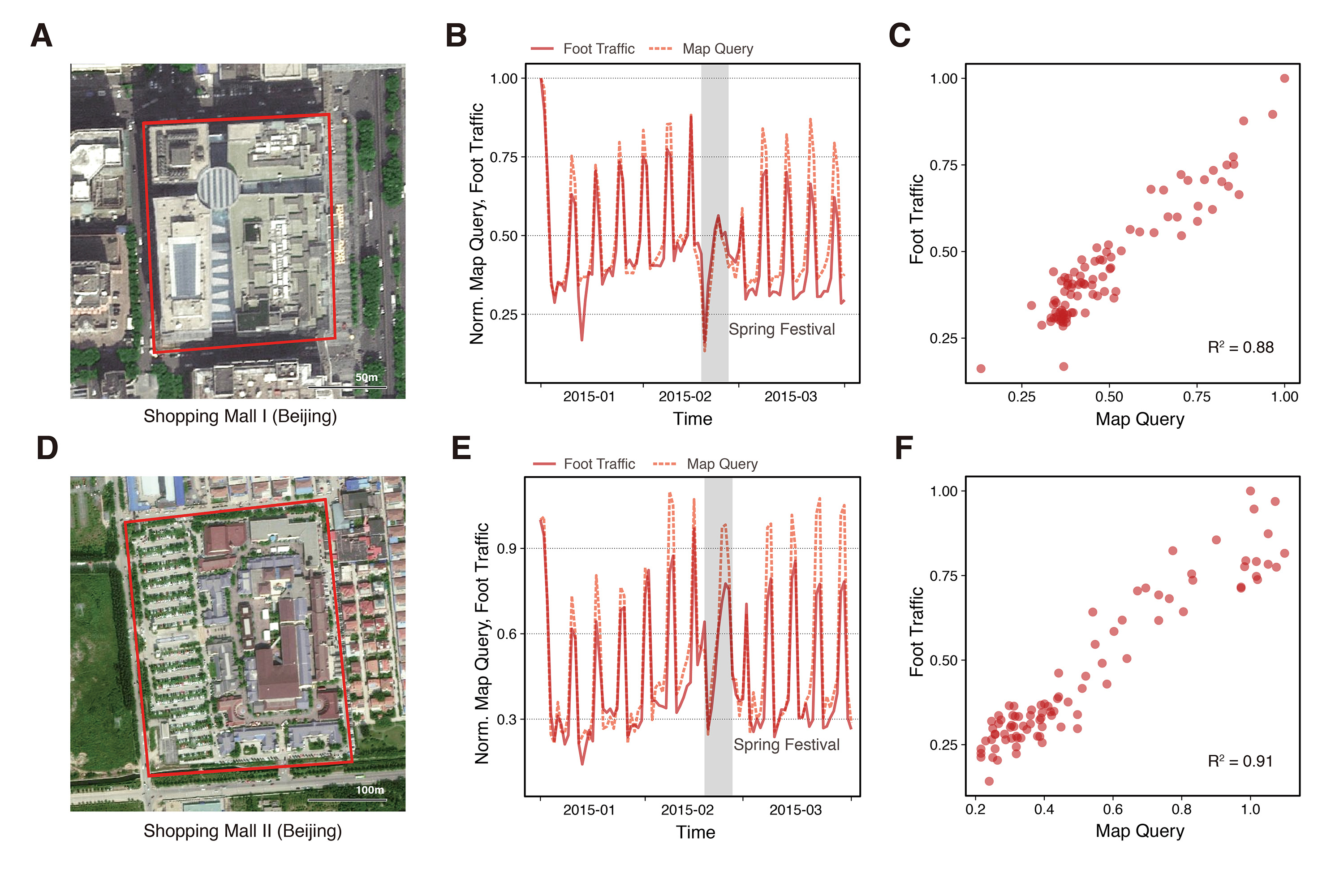}
\caption{Map queries and foot traffic. Spatial layout of Shopping Mall I (A) and II (D). The number of people searching for Shopping Mall I (B) and II (E) on Baidu Maps and the foot traffic of each place. Scatter plots of map queries and foot traffic of Shopping Mall I (C) and II(F). In both cases map query data is highly correlated with offline foot traffic, and both of them show cyclical patterns, peaking on weekends.}
\label{fig:figure6}
\end{figure}

\subsubsection*{2.2 Foot Traffic and Revenue Forecast}

We here demonstrate two cases to validate the potential power of map queries on revenue estimating and forecasting. The first one is revenue estimation and prediction of Apple Inc. (Apple) in China, and the second is box office `nowcasting' and fraud detection.

\begin{itemize}
\item \textbf{Apple sales in China}
\end{itemize}

As a listed company, Apple reports its earning results to the public quarterly with about one-month lag. According to the official numbers, Apple reported \$75.9b in revenue during the last quarter of 2015 (note that fiscal quarter is different with calendar quarter), and China, the second largest market around the globe, contributed \$18.4b. (In the same quarter a year prior, the total revenue was \$74.6b, and China market contributed \$16.1b.) In the fist quarter of 2016, Apple’s revenue (\$50.6b) was down year-over-year for the first time since 2003, and the revenue in China declined about 26\%.

Given that we are able to estimate the foot traffic volumes of offline stores through map queries, are they related to the reported revenue? Is it possible for us to predict the quarterly revenue ahead of the earnings’ announcements?

To answer these questions, we first select a list of flagship Apple Store in Mainland China (see the Supplementary Materials for the list), and then count the volume of map queries of all the stores. As shown in Figure 7A, the blue line spans from the last quarter of 2014 to the first quarter of 2015, the red line spans the corresponding time period of 2016. We also calculate the year-over-year changes of map queries (gray area in Figure 7A), and compare it with revenues (Figure 7B). As Figure 7 shows, the volume of map queries was up from a 15.4\% year-over-year growth in the last quarter of 2015, and down for a 24.5\% decline in the first quarter of 2016, comparing with a 14\% increase and 26\% decrease of revenues in the meantime. The impressively strong correlation indicates that map query data provides possibilities for us to `nowcast' the company’s revenues and reveal the future trends. Based on our analysis of latest data, we project that the Apple’s revenue in Greater China of second quarter of 2016 may decline between 23\% and 34\% on a year-over-year basis.

\begin{figure}[ht]
\centering
\includegraphics[width=\linewidth]{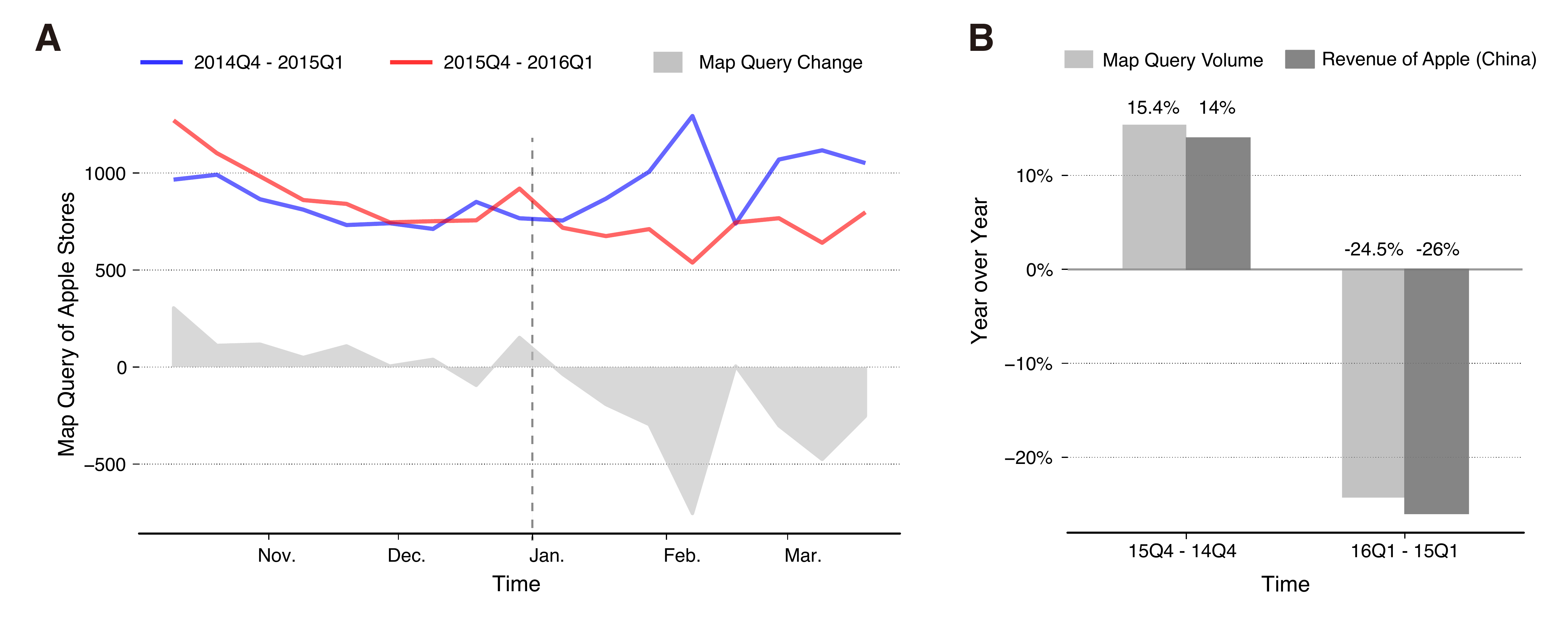}
\caption{Map queries of Apple Stores and revenues. (A) The volume of map queries of flagship Apple Stores in the mainland of China. Year over year changes of queries is plotted as gray areas. (B) The volume of map queries was up from a 15.4\% year over year growth in the fourth quarter of 2015, and down for a 24.5\% decline in the first quarter of 2016, comparing with a 14\% increase and 26\% decrease of revenues in the corresponding period. Based on our analysis of latest data, we project that the Apple’s revenue in Greater China of second quarter of 2016 may decline between 23\% and 34\% on a year-over-year basis. 
}
\label{fig:figure7}
\end{figure}

\begin{itemize}
\item \textbf{Movie box office `nowcasting' and fraud detection}
\end{itemize}

Going beyond estimation at firm level, we find that map query data is also predictive at specific consumer sector level. To better illustrate, we collect daily box office data over 2015 from China Box Office Database [58921.com], and then build models to `nowcast' daily box office by using map queries related to cinemas. Here we use the word `nocasting' instead of `forecasting', since we use map query to predict box-office of the same day, which is about one day earlier than official statistics. To reduce the effect of user search trends, we normalize the the volume of daily map query data and use it as independent variables in our regression models. Our baseline model is simple auto regression with 1 day and 7 day time lags as we have no further information about movies (e.g. production budgets, number of opened screens, which were used as input features in ref [\cite{goel2010predicting}]. The regression equations are similar to Eq.1 and Eq.2. 

As shown in Supplementary Table 1, map queries of cinema significantly improve the in-sample fitting, the $R^2$ of the baseline model is 0.489, while that number of map query model is 0.934 (see the Supplementary Materials for more regression results). To further investigate model's out-of-sample performance, we apply a rolling window forecasting similar to the method used in  [\cite{choi2012predicting, lazer2014parable}]. Given the selected date (July 1st, 2015 in this case), prediction in $t+1$ time step is based on estimates of all previous time periods ($\le t$).

The results are plotted in Figure 8, of which the gray line is the actual box office, blue line and red line denote the forecasting results of baseline model and our map query model, respectively. The absolute error of baseline model is almost always larger than map query model, and the mean absolute error (MAE) of map query model declines 67\% than baseline model with the value of 13.2 million RMB (the mean of daily box office is 118 million, the standard deviation is 75.8 million) as shown in Figure 8B.

What's more, we reveal some additional insights when analyzing these data of specific movies, and find that map queries are helpful in detecting box office fraud. For example, $Monster~Hunter$, China box office champ in 2015 with about 2.4 billion RMB revenue, was reported to artificially inflate its box office by schedule mid-night screenings in cinemas owned by movie producer with nearly no audience present [\cite{website:chinadaily}]. To check this, we select a list of suspected cinemas owned by the producer of $Monster~Hunter$ (about 30 cinemas), and a list of 30 other cinemas as a control group. We then calculate the volume of map queries of these locations from July 2015 to September 2015 when the movie was shown.

\begin{figure}[ht]
\centering
\includegraphics[width=0.7\textwidth]{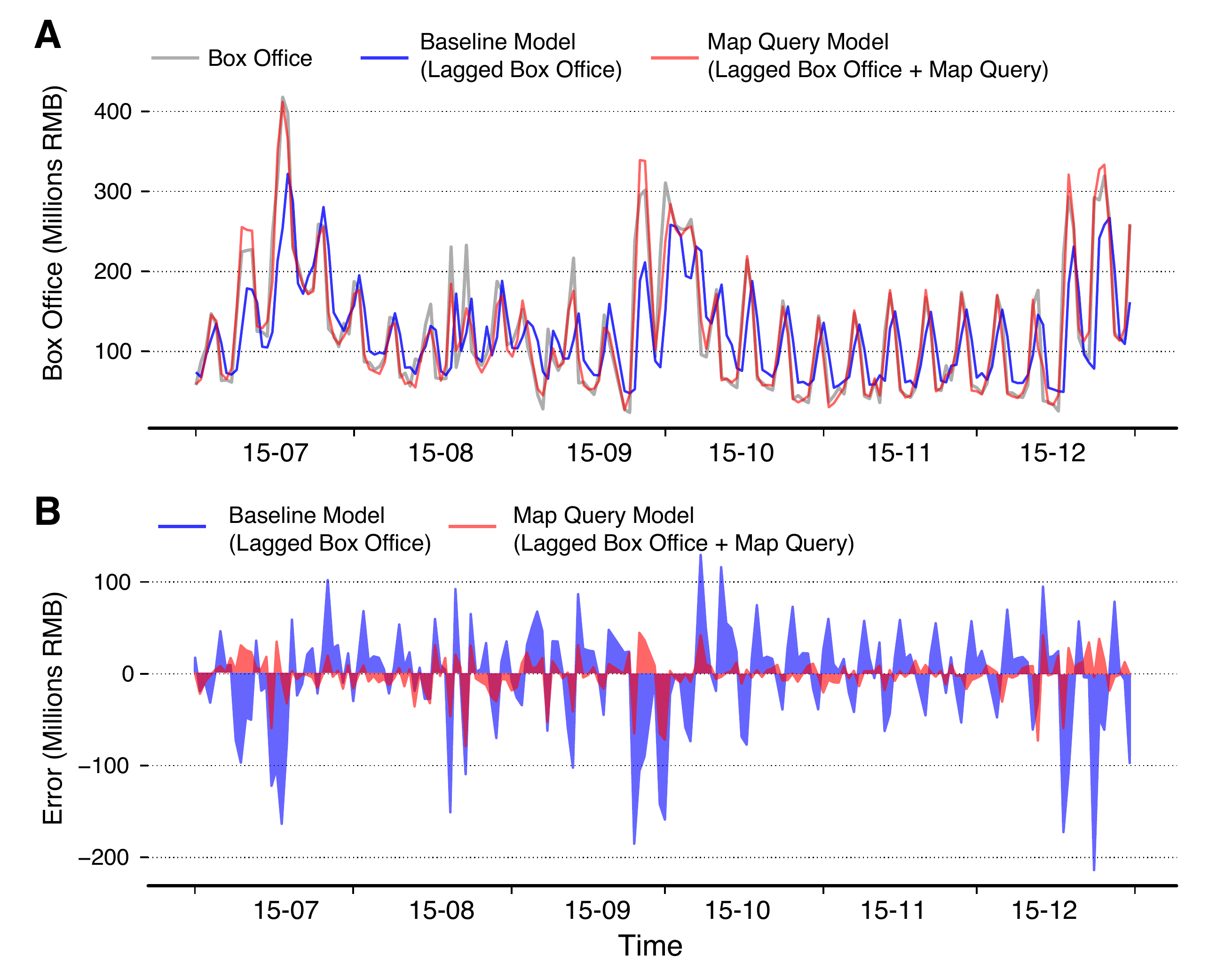}
\caption{Box office prediction. (A) Box office revenue (gray line), baseline model (blue line) and map query model (red line). (B) Predicting errors of baseline model (blue line) and map query model (red line). The absolute error of baseline model is almost always larger than that of map query model, and the MAE of map query model declined 67\% relative to baseline model.}
\label{fig:figure8}
\end{figure}

\begin{figure}[H]
\centering
\includegraphics[width=\linewidth]{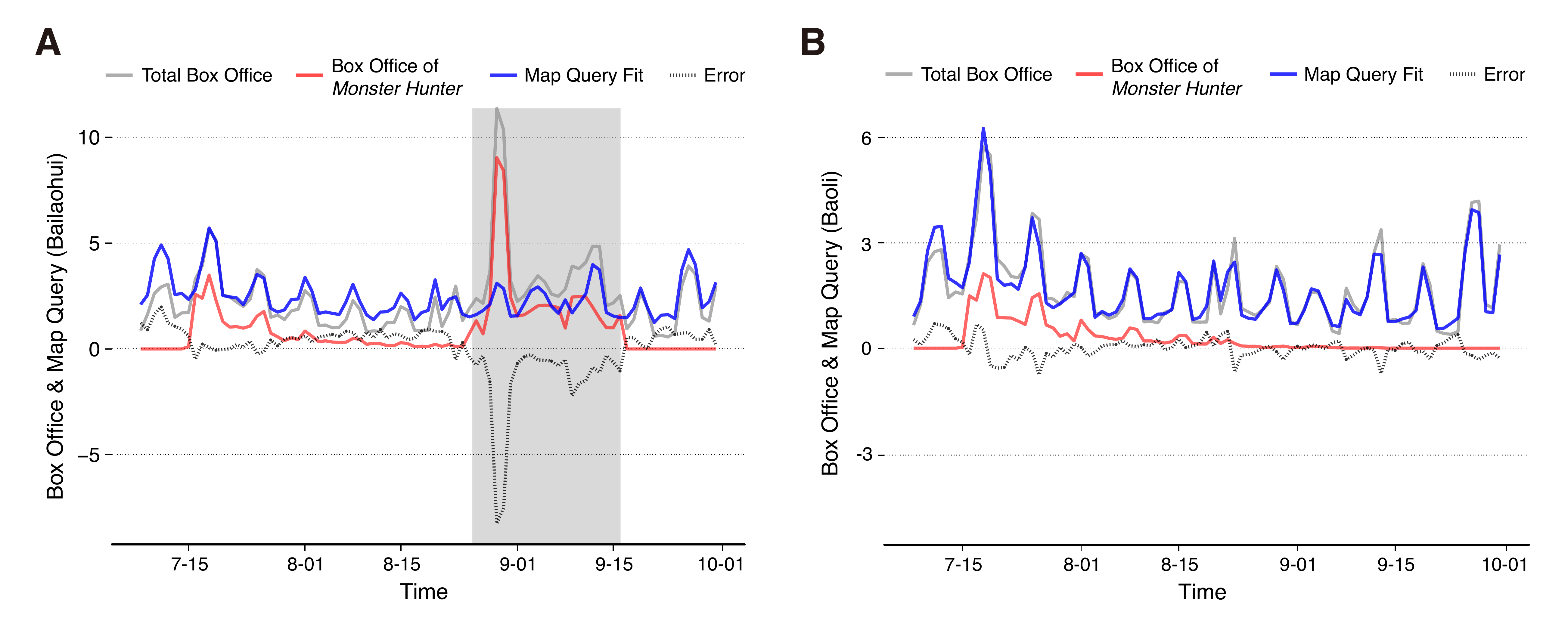}
\caption{Box office fraud detection. Movie $Monster~Hunter$ was suspected to artificially manipulate its box office by schedule mid-night screenings in cinemas owned by movie producer with nearly no audience present. We choose 30 POIs of these cinemas and set them as suspected group (A). We also pick another 30 cinema POIs as control group (B). The forecasting error of control group is small and stable, while in the suspected cinema set, there is a sharp increase of forecasting error from August 25 to September 16 (marked by gray box), such anomaly indicating high possibility of box office fraud.}
\label{fig:figure9}
\end{figure}

Figure 9A shows the results of suspected cinemas, and Figure 9B shows those of the control group. Gray lines and red lines are the total box office and $Monster~Hunter$ box office, respectively. Blue lines are the fitting results by map queries, and black dashed lines denote the forecast errors. By comparing the forecasting errors of two sets of selected cinemas, one can see that the forecasting error of control group is small and stable, while in the suspected cinema set, there is a sharp increase of forecasting error from August 25 to September 16 (marked by gray box), such anomaly indicating high possibility of box office cheating. 

\subsubsection*{2.3 Consumption Trends Index}

As demonstrated above, the volume of map queries is strongly correlated with foot traffic and is able to estimate offline consumer spending, and we therefore construct consumption indices of various industries by aggregating the map queries that belong to the same POI category. For example, one who searches a car dealership on Baidu Maps will be regarded as a potential consumer of automobile industry. In this paper, we present the consumption trends of four sectors, $Automobile~Sales$, $Restaurants$, $Finance\&Investment$, and $Tourism$, and compare our results with several other indicators. All indices except finance and investment sector are normalized by the mean value of year 2014 which is set to be 100.

\begin{figure}[H]
\centering
\includegraphics[width=\linewidth]{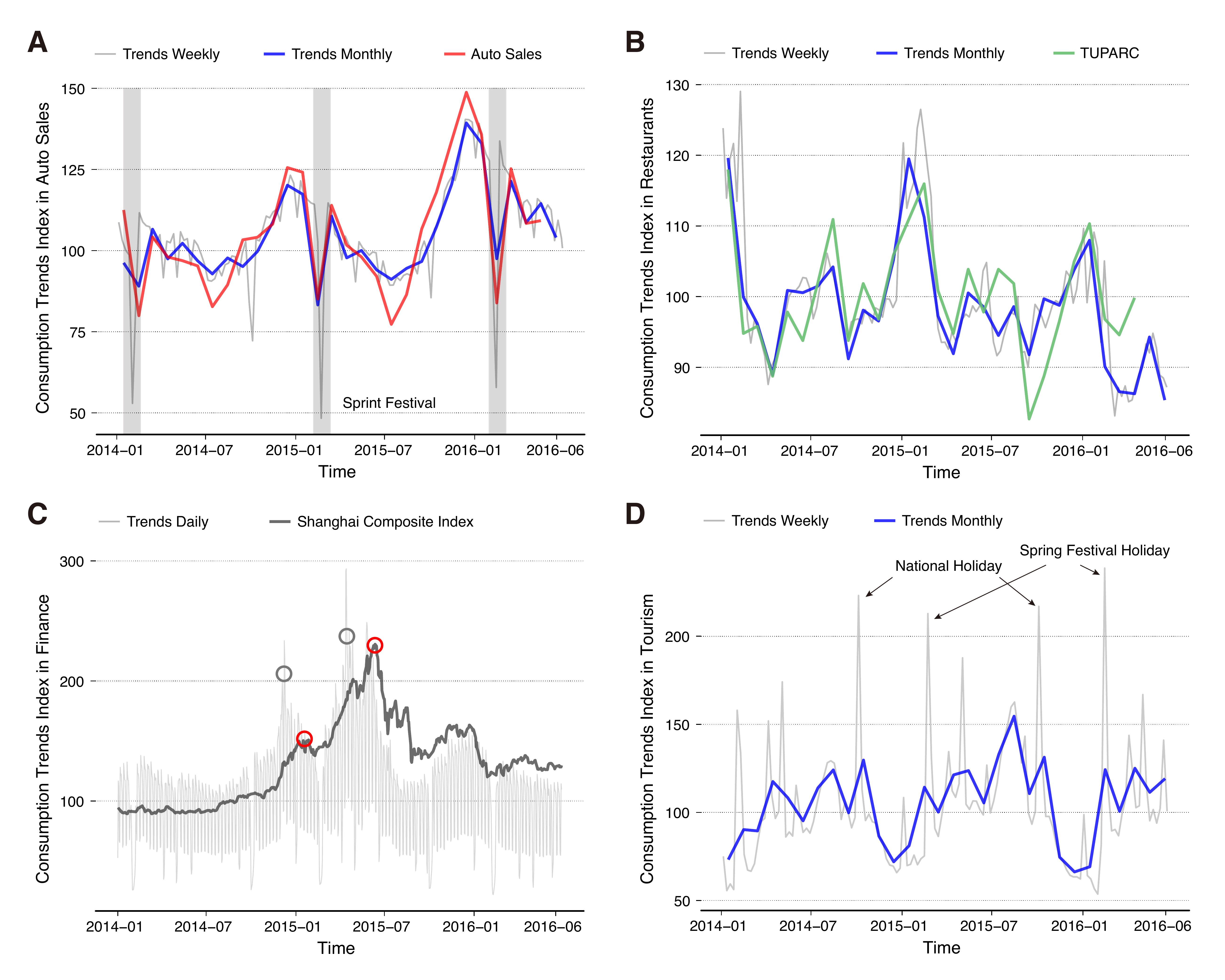}
\caption{Consumption Trends Index of various industries in service sectors. All indices except finance and investment sector are normalized by the mean value of year 2014 which is set to be 100. (A) Auto sales. (B) Restaurants. (C) Finance\&Investments. We mark the `turning point' of stock market and map query data by circles, which indicates that map query may be a leading indicator for stock market movements. (D) Tourism. Important holidays are labeled by arrows. The proposed Consumption Trends Indices of these sectors demonstrate that on a year-over-year basis, automobile sales are rising, restaurant spending is declining, finance investment is related to stock market volatility, and the travel spending is struggling with a moderate growth, all of which reflect the complicated reality of China’s service economy.}
\label{fig:figure10}
\end{figure}

Figure 10A shows the overall rising trend of automobile sales based on analysis of about 190,000 related locations, which is highly correlated with the sales data published by Automobile Association of China [http://www.caam.org.cn/]. The Pearson correlation is 0.913 with $P < 0.001$, indicating that the proposed trend could track the market changes effectively in a real-time fashion.

Figure 10B shows the relatively declining trend of restaurant sector including near 3 million restaurants. We compare our result with Tsinghua UnionPay Advisors Restaurant \& Catering (TUPARC) Index [\cite{website:unionpay}] which is a consumer spending indicator based on bank cards transaction data from top 100 restaurants ranked by China Cuisine Association. While it is consistent with TUPARC over most of the time period (the Pearson correlation coefficient is 0.764 with $p < 0.001$), our restaurant consumption trend started dipping since 2016 and fell new low of 85 in June.
%Such decline may partially result from the increasingly popularity of on-demand food delivery services in China.

Figure 10C shows the result of finance\&investment sector including 230,000 related locations. Interestingly, we find that it resembles the trends of Shanghai Composite Index. Our ongoing further investigation indicates that the proposed trend may be a leading indicator in stock market movement especially for 'turning points' (marked by circles). 
%Since map query reflects `consumer demand' from a bottom-up view, it could be no surprise to see the `price of good' changes with demand, and we leave further analysis in future works. 

Figure 10D shows the trend of travel sector comprising 300,000 tourist attractions. The overall trend grows moderately, and exhibits obvious spikes during major holidays in China such as Spring Festival, National Holiday and summer holiday that generally lasts from July to August. 

Our consumption trends of these sectors demonstrate that on a year-over-year basis, automobile sales are rising, restaurant spending is declining, finance investment is related to stock market volatility, the travel spending grows moderately, all of which reflect the complicated reality of China’s service economy.

\section*{Discussion}

In summary, we construct real-time indices to track the trends of employment and consumption in China at different scales via mobile big data. We employ several strategies to verify our assumptions and results. At national level, since there are no parallel statistics in government reports, we study several cases covered by media, and the numerical changes of employees of these cases could be tracked by our data and method successfully. At firm and sector level, we estimate the foot traffic by map queries, and it offers impressive results in predicting Apple's revenues in China and box-office earnings. We also construct indices to track the consumer spending trend of various industries in service sectors, and compare our results with other existing indicators.

Using mobile data to build economic indicators have great practical values. Compared with traditional survey based methods, mobile data is available nearly in real time, with larger sample size, finer resolution, and lower cost. Compared with social media data, mobile data is of higher coverage rate, more structured and robust [\cite{toole2015tracking}]. These features are quite valuable to market participants and policy makers who need timely and reliable data to make decisions. More importantly, the nature of mobility traces and location search data allows us to measure the individual-level behaviors in a much more direct fashion. We here introduce the concept of $Mobimetrics$: it dedicates to quantify the dynamics of social system by analyzing massive individual mobility data generated from smartphones, wearable devices, driverless cars and even Internet of Things in the near future with machine learning approaches, which may change the landscape of empirical research related to economics and social science.

Though we highlight the value of mobile data in measuring and predicting economic activities, we do not claim it could supplant official statistics. Especially, mobile data only covers people who use specific services, introducing potential bias to specific research topic. For example, the percentage of old people and children is lower in mobile data than real age structure in China (the comparison of the user's age and location distribution versus the surveyed results are shown in the Supplementary Materials), so if a study is designed to evaluate a policy on the elderly population, mobile data maybe not a good choice. In fact, the combination of survey and new data source could play their respective advantages, and achieve more accurate and abundant measurements.

This study is just the start application of $Mobimetrics$ and several improvements could be made in future research. First, in order to cover more sectors and build a more comprehensive economic measurement, more AOIs could be labeled in an automatic fashion by using machine learning and incorporating more date sources such as road network data and satellite images etc.. Second, the proposed economic indices could be further investigated through cross validation with more data sources. For example,  map queries of hospitals, which can be viewed as an important indicator for diseases monitoring, could be verified by data from the Centers for Disease Control and Prevention. Besides, we plan to combine more offline data sources, such as offline survey data, and transaction data to improve our methodology. Third, on-demand platform economy, or sharing economy, is now playing a more and more important role in our daily life, how to measure such economic activities via mobile data is a question worth exploring.

%%to be updated
All the proposed indices including Employment Index of industrial parks, Consumer Index of commercial areas, Consumption Trends Index will be published at the beginning of each month. 

\bibliography{reference}

\section*{Acknowledgments}

We thank all the members from Spatial Temporal Big Data Group of Big Data Lab (BDL) for helpful discussion.

\section*{Author contributions statement}
L.D. and H.W. designed the research and wrote the paper. L.D., S.C., Y.C. and Z.W. performed data analysis. C. L. helped with data pre-processing. All authors reviewed the manuscript. 

%\section*{Additional information}
% \textbf{Competing financial interests:} H.W., Y.C. and C.L. are full-time employees of Baidu. L.D., S.C., and Z.W. are interns at Baidu during this research period.

\begin{table*}[hb]
\centering
\begin{threeparttable}
\begin{tabular}{c c c cc c c c} 
\hline \hline
Variables & Shopping Mall I (1) & Shopping Mall I (2) & Shopping Mall II (1) & Shopping Mall II (2)\\ \hline
$y_{t-1}$ & 0.125 & 0.117** & 0.072 & 0.080* \\
& (0.068) & (0.038) & (0.062) & (0.034) \\
$y_{t-7}$ & 0.811*** & 0.078 & 0.866*** & 0.030 \\
& (0.068) & (0.072) & (0.062) & (0.076) \\
$Map~Query$ & & 2.60*** & & 5.10*** \\
& & (0.220) & & (0.413)\\
& & & &\\
Observations & 74 & 74 & 74 & 74\\
R-squared & 0.727 & 0.914 & 0.767 & 0.931\\
\hline \hline
\end{tabular}
\label{table:1}
\begin{tablenotes}
	\small
	\item \textbf{Note:} OLS regressions are estimated with a constant that is not reported in this table. Standard errors are shown in brackets. *** $p<0.001$, ** $p<0.01$, * $p<0.05$.
\end{tablenotes}
\caption{Regressions of foot traffic and map queries.}
\end{threeparttable}
\end{table*}

\end{document}